\documentclass{ws-procs9x6}
\begin{document}

\title{CONSTRUCTION OF THE DUAL GINZBURG-LANDAU THEORY FROM 
THE LATTICE QCD}

\author{H.~Suganuma, K.~Amemiya, H.~Ichie and Y.~Koma}

\address{Research Center for Nuclear Physics (RCNP), Osaka University, \\
Mihogaoka 10-1, Ibaraki, Osaka 567-0047, Japan}

\maketitle

\abstracts{ 
We briefly review the QCD physics and then 
introduce recent topics on the confinement physics.
In the maximally abelian (MA) gauge, 
the low-energy QCD is abelianized owing to 
the effective off-diagonal gluon mass 
$M_{\rm off} \simeq 1.2 {\rm GeV}$ induced by the MA gauge fixing.
We demonstrate the construction of the dual Ginzburg-Landau (DGL) 
theory from the low-energy QCD in the MA gauge 
in terms of the lattice QCD evidences on 
infrared abelian dominance and infrared monopole condensation. 
} 

\section{What is the Quark Confinement Physics ?}

About two centuries ago, Dalton and Avogadro introduced the idea of 
``atoms" and ``molecules" into the modern science. 
Since then, to find out the elementary object 
becomes one of the most important subject in the modern science. 

According to the experimental progress, the ``elementary object'' 
has been changed. The atom is divided into a nucleus and 
electrons, and the nucleus is divided into nucleons... 
Then, to separate out the more fundamental object becomes  
one of the most central philosophy in the particle physics. 

Nowadays, the nucleon is recognized to consist of quarks and gluons. 
Nevertheless, the individual quark cannot be 
separated from the nucleon, but is confined inside the nucleon. 
We can never observe the raw quark directly, 
and quarks are strongly bound each other.  
Then, the Quark Physic is a sort of many-body physics. 
This is the most different point from 
the other subjects in the elementary particle physics. 

Furthermore, such a curious quark feature is deeply related to 
the property of the QCD vacuum, which is not a large 
``empty box" but is regarded as the ``condensed matter" 
with a finite values of the gluon condensate and the quark condensate. 
Therefore, the Quark Physics requires the total sense of 
the particle physics, many-body physics and condensed matter physics.
(So, the Quark Physics looks like a jigsaw puzzle. 
We need individual precise knowledge keeping the total balance.)

\section{What is the Strategy of the Quark Physics ?}

To seek the elementary law is one of the most important aims 
in the particle physics. 
However, the aim of the Quark Physics is quite different, 
because we already know the elementary law of QCD. 
Surely, QCD tells us the ``rule" of the elementary interaction between 
quarks and gluons, but to solve QCD is another difficult problem.

\begin{figure}[hb]
\centering
\includegraphics[height=8cm]{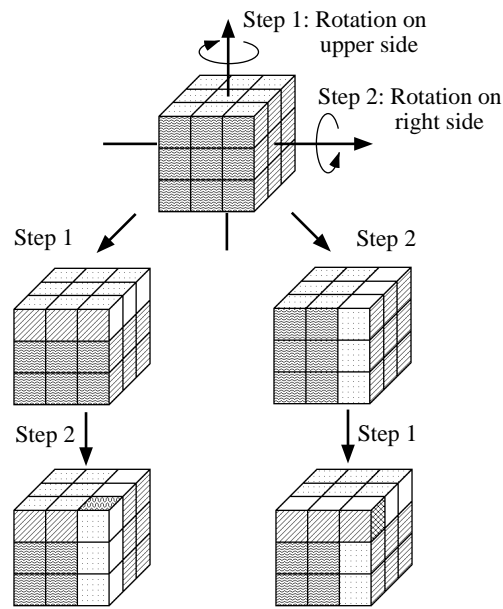}
\caption{
The difficulty and the interest of the Rubik cube originate from 
non-commutable operations based on the nonabelian nature 
of the rotational group. 
The configuration after Step 1 and Step 2 
depends on the order of these operations. 
Can you find a nonabelian nature of QCD in the Rubik cube ?}
\label{fig1}
\end{figure}

Then, the QCD physic is similar to the Rubik Cube ! 
The rule is simple, but to solve is rather difficult. 
This difficulty is also based on the non-commutable procedures  
of the ``nonabelian" rotational process as shown in Fig.1.
Here, a kind of ``locality" of the rotational process  
leads to variety and extremely large number of configurations, 
which makes this puzzle difficult and interesting. 
Thus, one may be able to find interesting analogy
between the Rubik cube and QCD, which is characterized 
by the nonabelian local symmetry. 

The present strategy for the Quark Physic seems to be 
the both-side attack from QCD and hadron phenomena. 
This is similar to the Nuclear Physics strategy. 
In fact, we already know the experimental data of the nuclear force 
between nucleons, which is the elementary law of nuclear dynamics. 
Nevertheless, we need several phenomenological 
models such as the liquid drop model, the shell model, the cluster model, 
the interacting boson model 
to understand the many aspects of the nuclear properties. 
(No nuclear physicist would say that nuclear physics is 
to solve the Schr\"odinger equation of nucleons.) 
Similarly in the nuclear physics, 
the linkage of the ``both sides", QCD and hadron phenomena, 
is one of the most important aims in the Quark Physics. 

\section{Why is the QCD Physics difficult ?}

The main difficulty of QCD originates from the nonabelian nature 
and the strong-coupling nature in the infrared region below 1 GeV. 
In the ultraviolet region, the QCD coupling becomes weak and then 
we can use the perturbation technique, where the nonabelian part can 
be treated as the perturbative interaction. 
But, in the infrared region, the nonabelian and 
the strong-coupling nature becomes significant. 

To see the difficulty on the nonabelian property of QCD, 
let us consider the simple electro-magnetic system. 
In the ordinary electro-magnetism, we can individually 
consider the partial electro-magnetic field 
formed by each charge, and the total electro-magnetic field can be 
obtained by adding these individual solutions. 
Here, additivity of the solution plays the key role, 
and this additivity originates from the linearity of the 
field equation, 
$
\partial_\mu F^{\mu\nu}=j^\nu, 
$
in the electro-magnetism. 

On the other hand, owing to the nonabelian nature, 
the QCD field equation becomes nonlinear as 
\begin{eqnarray}
\partial_\mu G^{\mu\nu} + ig [A_\mu, G^{\mu\nu}]=j^\nu, 
\end{eqnarray}
and then 
it is difficult to solve it even at the classical level, 
because the solution does not hold the additivity. 
So, we cannot divide the color-electromagnetic field into 
each part formed by each quark. Instead, the QCD system is 
to be analyzed as a whole system. 
(Furthermore, the QCD vacuum is the nontrivial medium 
with the gluon condensate.) 
For the analysis of QCD, the nonabelian nature provides a 
serious difficulty. 

\section{Abelianization of QCD -
Infrared Abelian Dominance in the Maximally Abelian Gauge}

The nonabelian nature is one of the characteristic features of QCD. 
However, by taking the maximally abelian (MA) gauge in QCD, 
one can make the nonabelian (off-diagonal) ingredients of QCD 
inactive for the infrared QCD properties such as  
quark confinement and chiral-symmetry breaking. 
We call these phenomena as the infrared abelian dominance 
in the MA gauge.$^{1-15}$ 

In the Euclidean QCD, the MA gauge 
is defined so as to minimize the total amount of off-diagonal gluons,  
$
R_{\rm off} \equiv \int d^4x \sum_{\mu, \alpha} |A_\mu^\alpha(x)|^2, 
$
by the SU($N_c$) gauge transformation. 
Here, $A_\mu^\alpha(x)$ denotes the off-diagonal gluon 
in the Cartan decomposition, 
$A_\mu= \vec A_\mu(x) \cdot \vec H+A_\mu^\alpha E^\alpha$. 
In the MA gauge, by removing the off-diagonal gluons, 
QCD can be well approximated as an abelian gauge theory 
like the electro-magnetism 
keeping the essence of the infrared QCD properties. 
This approximation is called as abelian projection. 

Owing to this remarkable feature of MA gauge fixing, 
the gluon field can be approximated to be abelian as 
$A_\mu^a(x)T^a \simeq \vec A_\mu(x) \cdot \vec H$ for the argument 
on long-distance physics. 
Accordingly, the field equation of the abelian-projected QCD 
becomes linear like the Maxwall equation,  
\begin{eqnarray}
\partial_\mu        F^{\mu\nu}= j^\nu, \quad 
\partial_\mu \tilde F^{\mu\nu}= k^\nu, 
\end{eqnarray}
with the color-electric current $j^\mu$ and the color-magnetic 
current $k^\mu$.
Thus, the additivity on color-electromagnetic fields $F^{\mu\nu}$
works in the abelian-projected QCD in the MA gauge. 
This is the most attractive point of the MA gauge. 

\section{Origin of Infrared Abelian Dominance - 
Mass Generation of Off-diagonal Gluons in the MA Gauge}

QCD in the MA gauge exhibits the following interesting properties. 
\begin{itemize}
\item 
In the MA gauge, the nonabelian SU($N_c$) gauge symmetry is 
reduced into the abelian U(1)$^{N_c-1}$ gauge symmetry 
with the global Weyl symmetry.$^{11}$
\item 
In the MA gauge, the off-diagonal gluon amplitude 
is strongly suppressed. 
As the result, there appears a strong randomness 
of the off-diagonal gluon phase in the MA gauge,  
which is found to be mathematical origin of abelian dominance 
for confinement.$^{10,12}$
\item 
As a remarkable fact, the color-magnetic monopole 
appears as the topological object in the MA gauge.$^{2,3,7,10,12}$ 
\item
Infrared abelian dominance holds for quark confinement 
and dynamical chiral-symmetry breaking in the MA gauge. 
\end{itemize}

As the physical origin of infrared abelian dominance, 
we recently find out effective mass generation of 
off-diagonal gluons in the MA gauge using the lattice QCD.$^{10,13}$ 
The effective off-diagonal gluons mass is measured 
as $M_{\rm off} \simeq$1.2 GeV 
from the behavior of the off-diagonal gluon propagator 
$G_{\mu \nu}^{\rm off}$ in the SU(2) lattice QCD 
as shown in Fig.2.
\begin{figure}[hb]
\centering
\includegraphics[height=4.5cm]{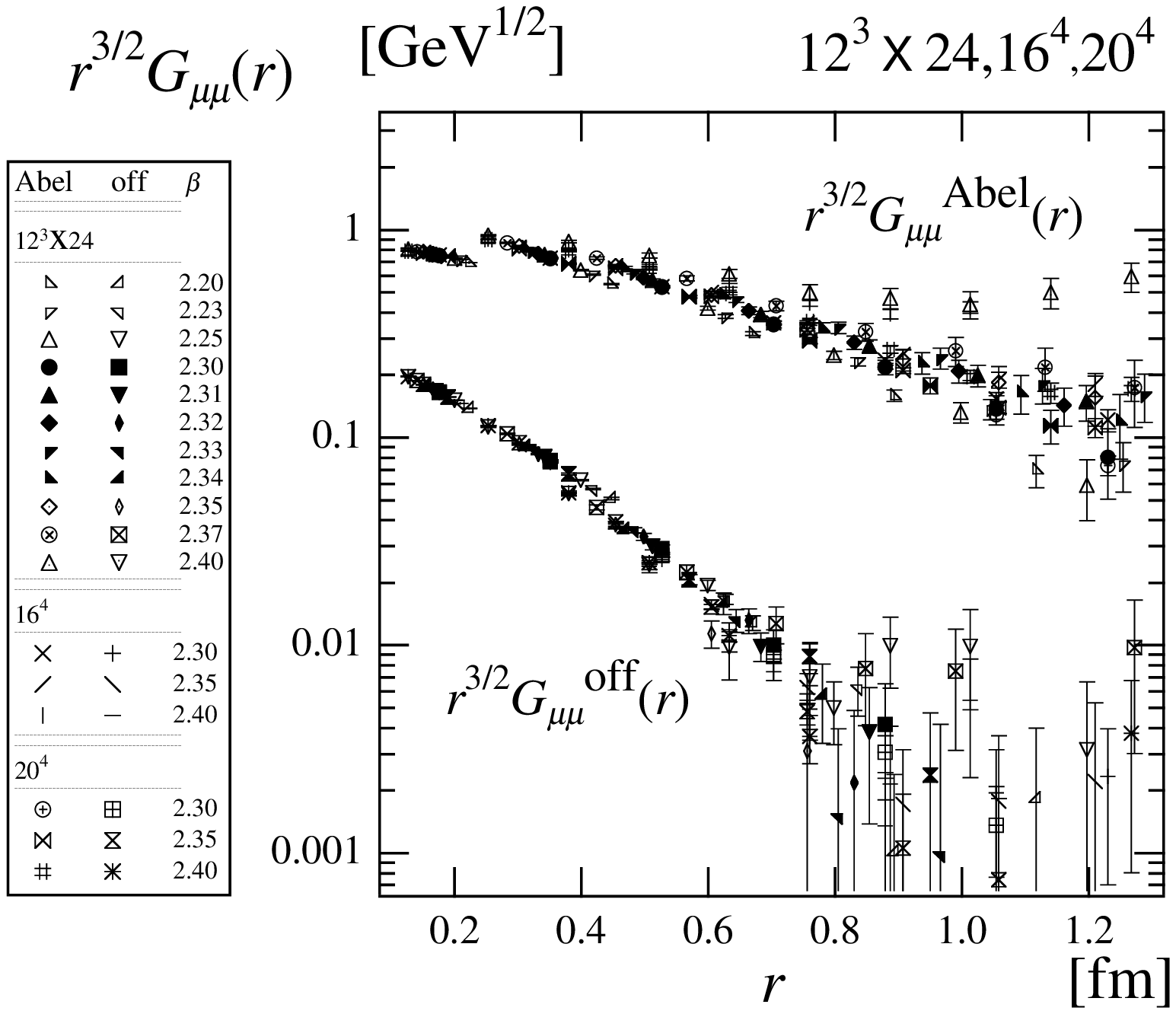}
\includegraphics[height=4.3cm]{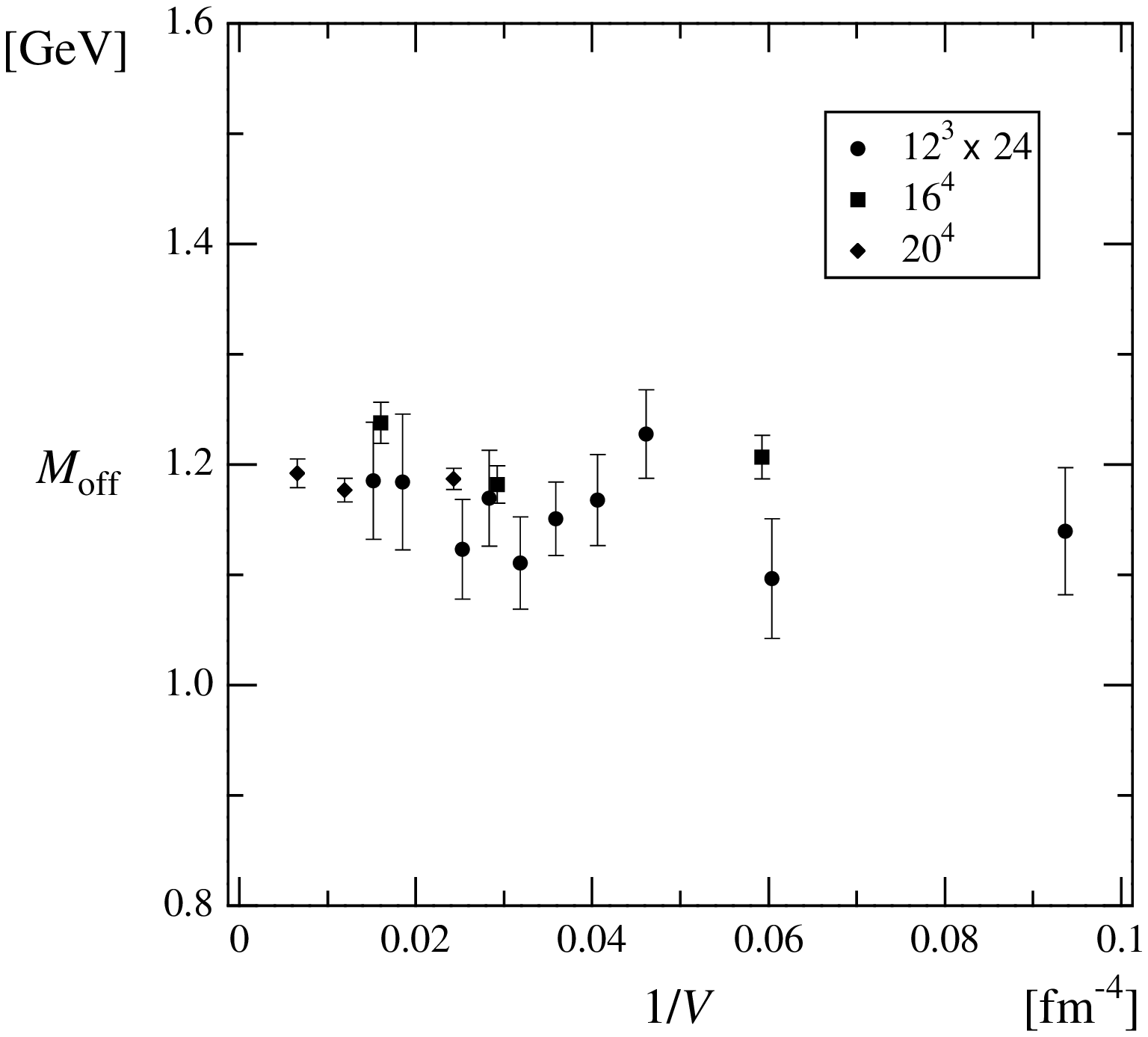}
\caption{
(a) The logarithmic plot of 
$r^{3/2} G_{\mu \mu}^{\rm off}(r)$ and 
$r^{3/2} G_{\mu \mu}^{\rm Abel}(r)$. 
The slope corresponds to the effective mass.
(b) The off-diagonal gluon mass $M_{\rm off}$ v.s. inverse lattice volume.}
\label{fig2}
\end{figure}

In the MA gauge, the off-diagonal gluon behaves as a massive field 
with a large mass about 1.2 GeV and then becomes inactive for the 
long-distance physics, 
similar to the negligible contribution of the massive weak boson 
to the long-distance force in the Weinberg-Salam model.
This is the essence of the infrared abelian dominance in the MA gauge. 

\section{Appearance of Color-Magnetic Monopoles 
in MA gauge in QCD}

In the MA gauge, the nonabelian SU($N_c$) gauge symmetry is 
reduced into the abelian U(1)$^{N_c-1}$ gauge symmetry. 
Owing to this partial gauge fixing, the color-magnetic monopole 
appears as the topological object corresponding to the 
nontrivial homotopy group, 
$\Pi_2({\rm SU}(N_c)/{\rm U(1)}^{N_c-1})=Z^{N_c-1}$, as was first 
pointed out by 't Hooft in 1980.$^{2}$ 

You may suspect the reality of the monopole in QCD. 
Then, do you know the following fact ? 
If the Weinberg-Salam model had simpler gauge symmetry SU(2)$_L$ 
instead of SU(2)$_L \times {\rm U(1)}$, this theory predicted 
the existence of the magnetic monopole with the mass about 
100 GeV ! 
So, the experimentalists might make much effort to create 
and to observe the magnetic monopole in order to get the Nobel prize !!
Unfortunately (?), the symmetry of the Weinberg-Salam model 
is SU(2)$_L \times$ U(1), and therefore this theory predicted 
the neutral current instead of magnetic monopoles. 

Grand Unified Theory (GUT) actually predicts the quite heavy 
magnetic monopole (GUT monopole) with the mass about $10^{16}$ GeV. 
The appearance of color-magnetic monopoles in the MA gauge in QCD 
has similar topological origin to the GUT monopole. 
Actually in the lattice QCD simulation in the MA gauge, 
there appears a global network of the monopole world-line 
covering the whole system as shown in Fig.3(a).$^{10,15}$ 
\begin{figure}[hb]
\centering
\includegraphics[height=4.5cm]{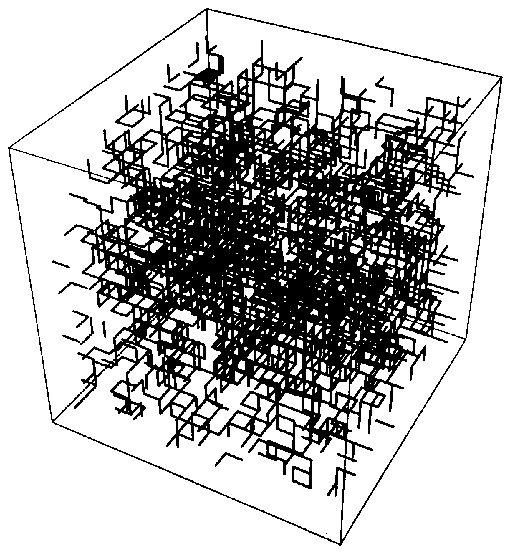}
\includegraphics[height=4.5cm]{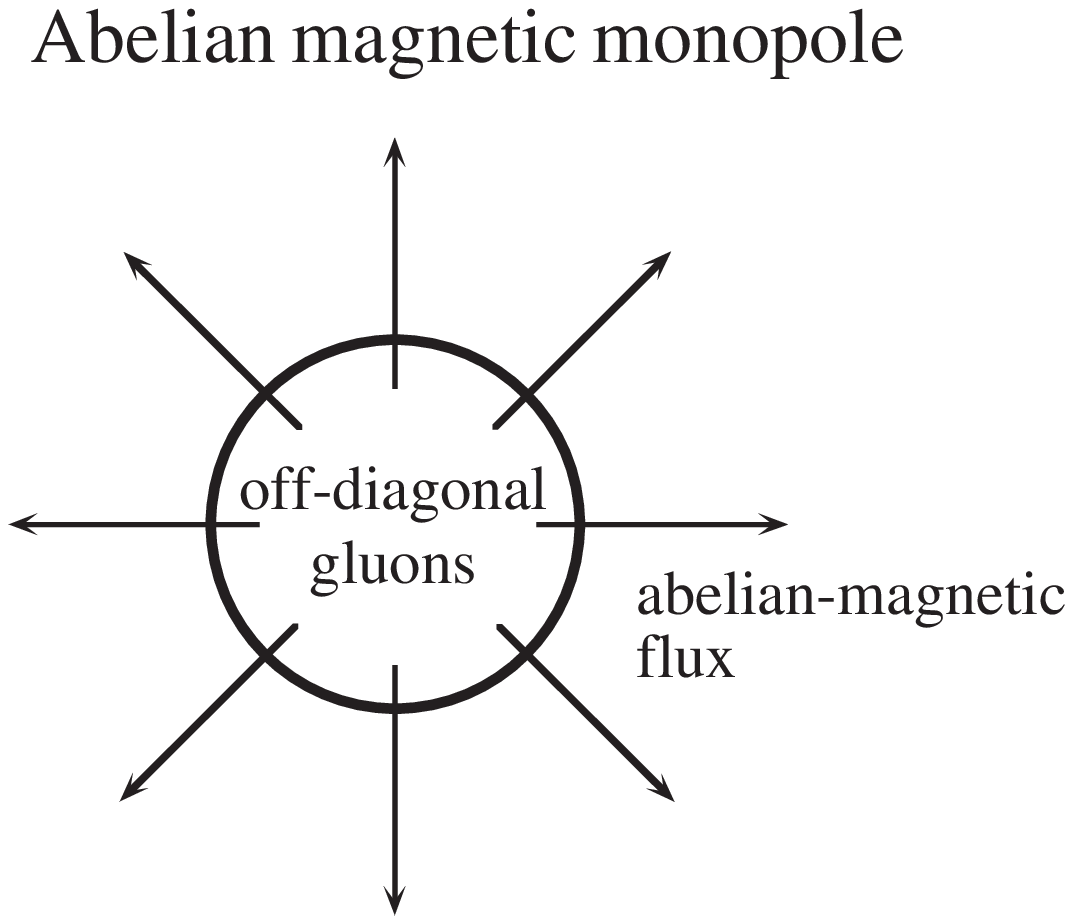}
\caption{
(a) The ${\bf R}^3$ projected world-line of the color-magnetic monopole 
in the MA gauge in the SU(2) lattice QCD 
with $16^3 \times 4$ and $\beta$=2.2 (confinement phase).
(b) The structure of the color-magnetic monopole. 
There remains large off-diagonal gluon component 
near the monopole center.}
\label{fig3}
\end{figure}

Here, even in the MA gauge, where the off-diagonal gluon element 
is strongly suppressed, around monopoles, 
there remains large off-diagonal gluon component.
This off-diagonal-gluon-rich region around the monopole provides 
an ``intrinsic size" and the structure of the monopole as shown in Fig.3(b), 
like the 't~Hooft-Polyakov monopole.$^{10,12,15}$ 
(The instanton is also an important topological object appearing in the 
nonabelian gauge manifold, and obviously the instanton needs 
full SU(2) components for existence. 
Therefore, instantons tend to appear in the off-diagonal-gluon-rich 
region around the monopole world-line in the MA gauge, which 
predicts the local correlation among monopoles, instantons and 
off-diagonal gluons.$^{10,15}$)

Using the lattice QCD, 
we find that the dual gluon $B_\mu$ is massive in the MA gauge,   
which is an numerical evidence of 
monopole condensation and the dual Higgs mechanism.$^{10,15}$ 
Then, quark confinement seems to be interpreted with monopole condensation, 
which was first proposed by Nambu in 1974.$^{1}$ 
Here, a color-singlet monopole appears as the dual Higgs particle,$^{11}$ 
and its fluctuation can be observed as a scalar glueball with $J^{CP}=0^{++}$. 

In fact, the dual superconductor scenario for quark confinement  
predicts a massive scalar glueball, 
like the Higgs particle in the Weinberg-Salam model.

\section{Summary and Outlook}

To conclude, the lattice QCD in the MA gauge exhibits 
{\it infrared abelian dominance} and 
{\it infrared monopole condensation}, 
and therefore the dual Ginzburg-Landau (DGL) theory$^{3-15}$
can be constructed as the infrared effective theory 
directly based on QCD in the MA gauge.

Historically, an interesting interpretation of phenomena has eventually lead 
to new finding on the particle as follows.
\begin{itemize}
\item
To interpret the origin of strong nuclear force,   
Yukawa predicted existence of mesons.
\item
Nambu and Goldstone proved that 
spontaneous chiral-symmetry breaking
leads to inevitable existence of light pseudoscalar particles (pions). 
\item
From the classification of hadrons, 
Gell-Mann and Zweig found quarks.  
\item
In near future, from further studies on the quark-confinement mechanism, 
new important finding may be done 
on the existence or the feature of the monopole (a scalar glueball). 
\end{itemize}

\section*{Acknowledgments}
We would like to thank Professor Yoichiro Nambu 
for his useful suggestions.
We acknowledge also Professors Hiroyasu Ejiri 
and Hiroshi Toki for their continuous warm encouragements at RCNP, 
Osaka University.

\begin{figure}[hb]
\centering
\includegraphics[height=15.5cm]{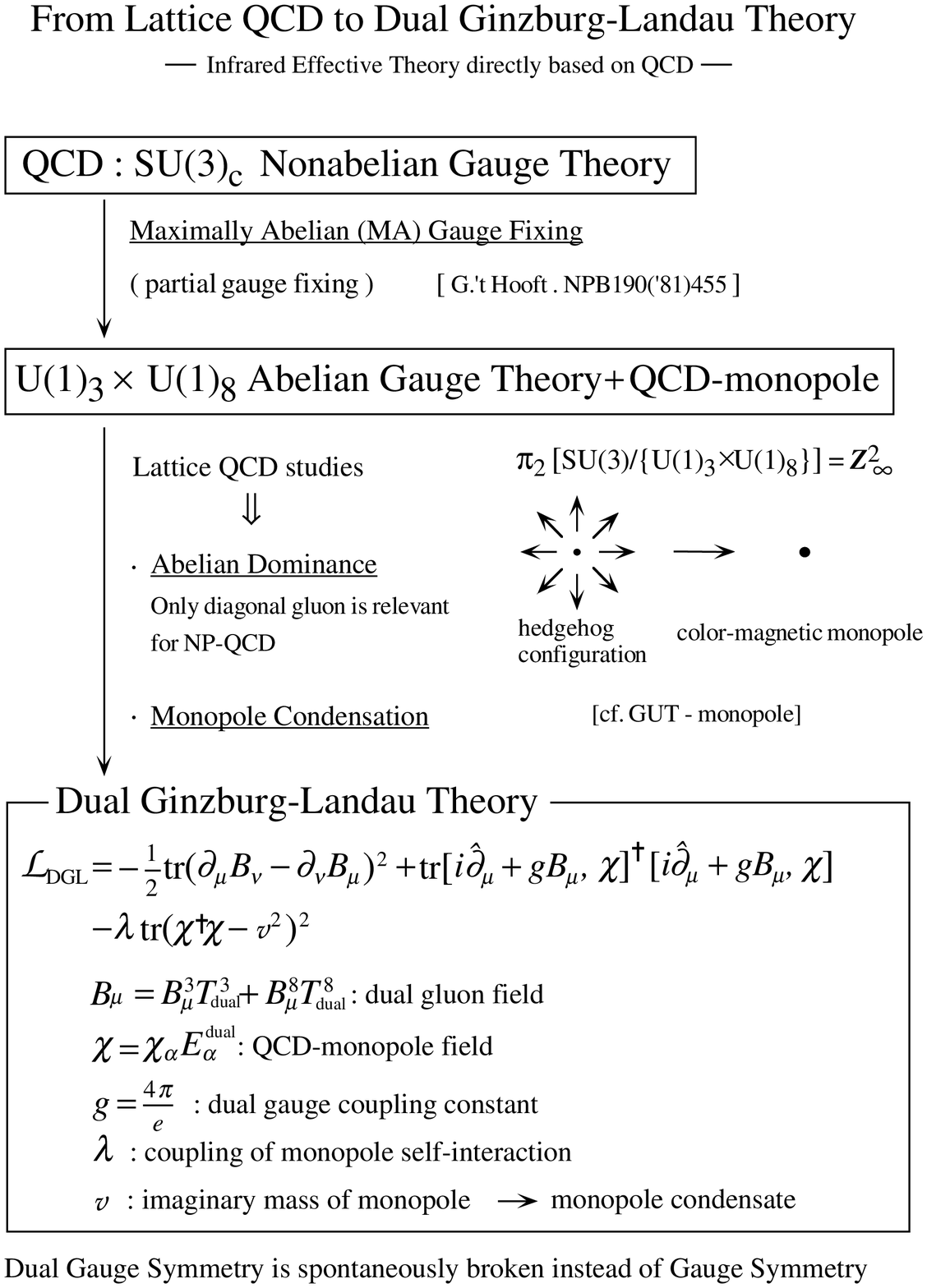}
\caption{
Construction of the dual Ginzburg-Landau (DGL) theory 
from the lattice QCD in the maximally abelian (MA) gauge.}
\label{fig4}
\end{figure}

\end{document}